\documentclass[12pt]{article}
\usepackage{amsmath}
\usepackage{epsfig}

\def\pw#1{^{{#1}}}
\def\eqref#1{(\ref{#1})}
\def\eqrefm#1#2{(\ref{#1}--\ref{#2})}
\def\cca{complex conjugate amplitude}
\def\tk{\tilde\kappa}
\def\tf{\tilde{f}}
\def\tV{\tilde{V}}
\def\tu{\tilde{u}}
\def\tv{\tilde{v}}
\def\xu{\underline{x}}

\def\epsu{\underline{\eps}}
\def\k2{\underline{k}}
\def\p2{\underline{p}}
\def\q2{\underline{q}}
\def\T2{\underline{T}}
\def\U2{\underline{U}}
\def\Q2{\underline{Q}}
\def\V2{\underline{V}}
\def\B2{\underline{B}}

\def\beeq{\begin{eqnarray}} \def\eeeq{\end{eqnarray}}
\newcounter{hran} 
\renewcommand{\thehran}{\arabic{hran}}

\def\bmini{\setcounter{hran}{\value{equation}}
  \refstepcounter{hran}\setcounter{equation}{0}
  \renewcommand{\theequation}{\thehran\alph{equation}}\begin{eqnarray}}

\def\bminiG#1{\setcounter{hran}{\value{equation}}
\refstepcounter{hran}\setcounter{equation}{-1}
\renewcommand{\theequation}{\thehran\alph{equation}}
\refstepcounter{equation}\label{#1}\begin{eqnarray}}

\def\emini{\end{eqnarray}\relax\setcounter{equation}{\value{hran}}\renewcommand{\theequation}{\arabic{equation}}}

\def\refup#1{~$\pw{\mbox{\small #1}}$}

\def\fun#1#2{\lower3.6pt\vbox{\baselineskip0pt\lineskip.9pt
  \ialign{$\mathsurround=0pt#1\hfil##\hfil$\crcr#2\crcr\sim\crcr}}}

\def\eV{{\rm e\kern-0.12em V}}

\def\eps{\epsilon}

\def \al {\relax\ifmmode{\alpha}\else{$\alpha${ }}\fi}
    \def\Re{\mathop{\rm Re}}
\def\abs#1{\left| #1\right|}

\def\ben{\begin{enumerate}}  \def\een{\end{enumerate}}
\def\bit{\begin{itemize}}    \def\eit{\end{itemize}}
\def\beq{\begin{equation}}   \def\eeq{\end{equation}}
\def\bea{\begin{eqnarray}}  \def\eea{\end{eqnarray}}

\def\kp{\relax\ifmmode{k_\perp}\else{$k_\perp${ }}\fi}
\def\kps{\relax\ifmmode{k_\perp\pw2}\else{$k_\perp\pw2${ }}\fi}
\def \as{\relax\ifmmode\alpha_s\else{$\alpha_s${ }}\fi}

  \def\refup#1{~$\pw{\mbox{\scriptsize\cite{#1}}}$}
  \def\refupd#1#2{~$\pw{\mbox{\scriptsize\citd{#1}{#2}}}$}
   \def\citd#1#2{\cite{#1,#2}}

\def\np#1#2#3{{\em Nucl.~Phys.}~\underline{B#1} (19#3) #2}
\def\pl#1#2#3{{\em Phys.~Lett.}~\underline{#1B} (19#3) #2}

\def\pr#1#2#3{{\em Phys.~Rev.}~\underline{#1} (19#3) #2}

\def\prl#1#2#3{{\em Phys.Rev.Lett.}~\underline{#1} (19#3) #2}

\begin{document}

\thispagestyle{plain}
\setcounter{page}{1}
\vbox to 1 truecm {}

\begin{flushright}
April 1998 \\[0.1cm]

BI-TP 98-06\\
IFUM 617/FT\\
CU-TP-886\\
LPTHE-Orsay 98-19
 \\
\end{flushright}

\vfill
\def\cen{\centerline}
\renewcommand{\thefootnote}{\fnsymbol{footnote}}

\cen{{\bf\large     MEDIUM-INDUCED RADIATIVE ENERGY LOSS; }}
\cen{{\bf  \large   EQUIVALENCE BETWEEN THE BDMPS }}
\cen{{\bf  \large   AND ZAKHAROV FORMALISMS\footnote{
This research is supported in part by the US Department of Energy
under GRANT DE-FG02-94ER40819.}
 }}

\vskip 1.5 truecm
\centerline
{\bf R.~Baier~$\pw1$, Yu.~L.~Dokshitzer~$\pw2$, A.~H.~Mueller~$\pw3$
 and D.~Schiff~$\pw4$}
\vskip 10 pt
\centerline{{\it $\pw{1 }$Fakult\"at f\"ur Physik,
Universit\"at Bielefeld, D-33501 Bielefeld, Germany}}
\centerline{{\it $\pw{2 }$ INFN sezione di Milano, via G. Celoria 16,  
20133 Milan, Italy\footnote[2]
{Permanent address: Petersburg Nuclear Physics Institute, Gatchina, 
 188350 St. Petersburg, Russia } }}
\centerline{{\it $\pw{3 }$Physics Department, Columbia University, 
New York, NY 10027, USA}}
\centerline{{\it $\pw{4 }$
LPTHE\footnote[3] { Laboratoire associ\'e au
Centre National de la Recherche Scientifique}
, Universit\'e  Paris-Sud, B\^atiment 211, F-91405 Orsay, France}}
 
\renewcommand{\thefootnote}{\arabic{footnote}}
\vskip 2 cm
\cen{\bf Abstract}
\vskip 5pt
\noindent
{We extend the BDMPS formalism for calculating radiative energy loss
  to the case when the radiated gluon carries a finite fraction of the 
  quark momentum. 
Some virtual terms, previously overlooked, are now included. The
equivalence between the formalism of BDMPS and that of B.~Zakharov is
explicitly demonstrated.}
\vfill \eject

\section{Introduction}

Over the past few years there has been much work done$^{\mbox{\scriptsize{[1--6]}}}$
studying the radiative energy loss of high energy partons passing
through hot and cold matter. These studies are extensions to QCD of
the analogous QED problem considered long ago by Landau, Pomeranchuk
and Migdal\refupd{r7}{r8}.
In one version of this work initiated in Refs.~\cite{r1} and \cite{r2}  
and continued in Refs.~\cite{r3} and \cite{r4} one follows the
multiple scattering\refupd{r9}{r10} of the high energy parton in the
QCD matter and the radiative gluon spectrum induced by the multiple
scattering is evaluated. A number of interesting, and perhaps
surprising results were found. The energy loss of a high energy jet in 
a hot QCD plasma appears to be much larger than that in cold nuclear
matter even at moderate, say 200 MeV, temperatures of the plasma. When 
a very high energy parton passes through a length $L$ of hot or cold
matter the induced radiative energy loss is proportional to $L^2$. 
A curious relation, $-dE/dz=\mbox{const}\cdot{\as N_c}p_{\perp W}^2$, was found 
between the energy loss and the width of the Gaussian transverse
momentum broadening of a high energy parton in QCD matter.

A different and very elegant approach to the energy loss problem has
been developed by Bronislav Zakharov\refupd{r5}{r6}. 
In his approach the
gluon radiation probability is given by the difference of the
probabilities of a virtual fluctuation of a high energy quark into a
quark and a gluon occuring in the vacuum and in the medium. The
formalism uses a clever device of treating the quark in the
\cca\ as an antiquark in the amplitude. What is
finally calculated then is the amplitude for a quark-antiquark-gluon
system to pass through a QCD medium without inducing inelastic
reactions in the medium. 
For a medium having many scatterers such an amplitude is not small
only if the quark-antiquark-gluon system is compact in transverse
coordinate-space, and this compactness makes the process
perturbatively calculable in QCD. 

One of the major purposes of the present work is to show that the
BDMPS\refupd{r3}{r4} and Z\refupd{r5}{r6} formalisms are equivalent.
However, before showing this equivalence it is necessary to extend the 
BDMPS formalism beyond the soft gluon approximation where it was
originally formulated. 
It is also necessary to include some virtual graphs which were omitted 
in our original formulation and which led to some numerical
discrepancies when our results were compared to those 
in Refs.~\cite{r5} and \cite{r6}.

In Secs. 2 and 3 we outline a derivation of the BDMPS formulas for the 
radiative gluon spectrum for a high energy quark passing through
either a hot or cold QCD medium, which formulas are valid even when
the radiated gluon carries a finite fraction of the quark's momentum.
Special care is taken to show which factors have changed from our
previous results when all terms corresponding to virtual corrections
(elastic scattering in the medium) are included. 

In Sec. 4 we evaluate our formulas for the radiative gluon energy spectrum. 
We consider both the case when the quark approaches the medium from
outside and when the quark is produced by a hard scattering in a medium. 

In Sec. 5 we show the equivalence between our approach and that of
B.~Zakharov.  We show that our formula \eqref{eq:30} for the radiative
spectrum leads to \eqref{eq:51} which is equivalent to (4) of 
Ref.~\cite{r6}. 
We then explain more qualitatively why the two formalisms, seemingly
very different, are in fact completely equivalent.

\section{The Born term for the amplitude}
In this section we calculate the lowest order terms for the emission
of a gluon from a quark which may enter the QCD matter from the vacuum 
or which may be produced in the matter through a hard interaction. We
begin by giving the basic vertex for emission of a gluon of momentum
$k$ from a quark of momentum $p$. 

We shall continue to call the
fermion from which the gluon is emitted a quark. However, in our final 
formulas results for spin 1/2 fermions of any colour representation,
$R$, can be obtained by multiplying the gluon emission spectrum 
\eqref{eq:39b} 
by the ratio of the ``colour charges'', $C_R/C_F$.
The generalisation to spin--1 (gluon) and spin--0 
projectiles will also be given in Sec.~\ref{sec:specandloss}.

\subsection{The basic vertex}
The basic vertex for gluon emission 
is given by
\begin{equation}
  \label{eq:1}
\mbox{
\epsfig{file=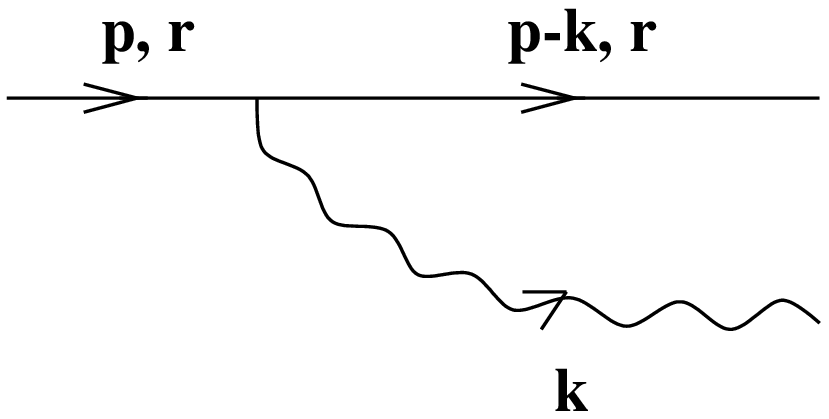,width=4cm}}
\>=\quad
\bar{u}_{r'}(p-k)\>\gamma\cdot\eps\> u_r(p) 
= \frac{2\delta_{rr'}}{x\sqrt{1-x}} \epsu\cdot\{\k2-x\p2\}  \>.
\end{equation}
We use a notation $\{\>\}$ such that for any two-dimensional vector
$v_i$, $i=1,2$ 
\begin{equation}
  \label{eq:2}
  \{v_i\} = \left(1-\frac{x}{2}\right)v_i - i\,\frac{x}{2}\, \eps_{ij}\, v_j\>,
\end{equation}
where $\eps_{ij}$ is the antisymmetric tensor; 
$\eps_{12}=-\eps_{21}=1$, $\eps_{11}=\eps_{22}=0$. 
The gluon polarisation vector is
\begin{equation}
  \label{eq:3}
  \eps_{\mu} = (\eps_0,\eps_z,\epsu) = \left( \frac{\epsu\cdot\k2}{2k}, 
           -\frac{\epsu\cdot\k2}{2k}, \epsu\right).     
\end{equation}
In \eqrefm{eq:1}{eq:3} we assume that 
\bminiG{eq:4}
\label{eq:4a}
\abs{\k2} &\ll& \abs{\,\vec{k}\,}=k\>, \\
\label{eq:4b}
\abs{\p2} &\ll& \abs{\,\vec{p}\,}=p\>, 
\emini
but we do not assume that $x=k/p$ is small.
The vertex is diagonal in the quark spin indices $r$ and $r'$ because of the 
vector nature of the coupling and because of our assumption that the
fermion is massless. 

\subsection{The Born terms for the amplitude; scattering in the medium}

In our procedure one explicitly integrates the emission time, $t$, of
the gluon between the times the quark, or the quark-gluon system, 
scatters in the medium. Consider, for example, the graphs in Fig~2a
where the quark scatters inelastically in the medium at time $t_1$ and
then later emits the gluon at time $t$. The $t$-dependent phase factor 
associated with this graph is
\begin{equation}
  \label{eq:5}
  \exp\left[\, i t\left(\frac{\k2^2}{2k}+ \frac{(\p2-\k2)^2}{2(p-k)}
-\frac{\p2^2}{2p}\right)\right]
\>=\> \exp\left[\, i t\frac{(\k2-x\p2)^2}{2x(1-x)p}\right]\>.
\end{equation}
The lower limit of the $t$-integration is $t_1$ while the upper limit 
will be the time of the next interaction or $t=\infty$ in case the
quark-gluon system leaves the medium without further interactions. 
Keeping the lower limit of the integral of the expression \eqref{eq:5}
and multiplying by the gluon emission vertex gives the factor
\begin{equation}
  \label{eq:6}
  4\,ip\sqrt{1-x}\>\exp\left[\,i
    t_1\frac{(\k2-x\p2)^2}{2x(1-x)p}\,\right]
\T2_a\cdot\epsu\>,
\end{equation}
where
\bminiG{eq:7}
  \label{eq:7a}
  \T2_a &=& \frac{\{\k2-x\p2\}}{(\k2-x\p2)^2}\>.
\eeeq
Similarly one can evaluate the emission term and energy denominators
in the remaining graphs b--g of Fig.~1
to obtain
\beeq
\T2_b &=& -\frac{\{\k2-x\p2-(1-x)\q2\}}{(\k2-x\p2-(1-x)\q2)^2} \>, \\
\T2_c &=& -\frac{\{\k2-x\p2+x\q2\}}{(\k2-x\p2+x\q2)^2} \>, \\
\T2_d &=& -\T2_e\>=\> -\T2_f\>=\> \T2_a \>, \\
\T2_g &=& -\frac{\{\k2-x\p2 -\q2\}}{(\k2-x\p2 -\q2)^2} \>.
\emini
The $(-1)$ factors in some terms in \eqref{eq:7} occur when $t_1$ is
the upper limit of the $t$ integration, as happens for the graphs
b, c, e, f and g of Fig.~1.

Graphs a--c of Fig.~1
correspond to inelastic reactions with the medium while graphs d--g
correspond to forward elastic scatterings in the medium. 
For terms a--c there are corresponding inelastic reactions in the
complex conjugate gluon emission amplitude. 
In the approximation that the forward elastic amplitude for quark
scattering off particles in the medium is purely imaginary, the elastic 
and inelastic terms are proportional to the same function $V$ to be
defined more precisely below. 

It is remarkable that the gluon emission vertices 
are proportional to exactly the same combinations of transverse momenta 
whose squares enter in 
the energy denominators for each of the terms in \eqref{eq:7}. 
This is a consequence of gauge invariance. 
The fact that the emission
amplitude is inversely proportional to the {\em first}\/ power of the relevant
transverse momentum follows from the Gribov bremsstrahlung
theorem\refup{rG}, the generalisation of the Low-Barnett-Kroll 
theorem\refup{rLBK}
to hard collinear radiation, $x\sim1$, $k_\perp\ll p$. 
This will prove crucial in obtaining a simple form for the gluon
emission spectrum when $x$ is not small. 

It is our convention that the time of emission of the gluon in the
complex conjugate amplitude, $t'$, is later than the emission time in
the amplitude. 
The opposite sequence of times will be accounted for by a factor of 2 in
the gluon spectrum \eqref{eq:28}.
This convention means that for the terms in the complex conjugate
amplitude corresponding to a--c of Fig.~1
it is only the quark and not the quark-gluon system which scatters
inelastically off the medium at time $t_1$. Thus the colour factors
associated with the scattering in the medium for the graph in Fig.~1a
is obtained by considering the graph of Fig.~2a
where the part of the graph to the right of the vertical line (cut) is
the complex conjugate amplitude. The initial and final $p$-lines have
the same colour. The graph in Fig.~2b
corresponds to the graph in Fig.~1b
along with the corresponding complex conjugate amplitude. Using the
formulas for colour factors given in Appendix~B of Ref.~\cite{r3} it
is straightforward to evaluate the ``colour factors'' associated with 
the hooking of the $q$-lines in Fig.~1
(and Fig.~2)
with the quark or quark-gluon passing through the medium, namely
\bminiG{eq:8}
F_a &=& -2\,F_d \>=\> -2\,F_f \>=\> C_F\>, \\
F_b &=& -F_e \>=\> F_g \>=\> \frac{N_c}{2}\>, \\
F_c &=& \frac{-1}{2N_c} \>.
\emini
In addition to pure colour factors the $F$'s also include a factor of
$(-1)$ for virtual terms and a factor of $1/2$ for virtual terms where
the $q$ and $-q$ lines attach to the same quark or gluon line. 
These factors naturally appear in a Feynman diagram description of
Glauber multiple scattering. For quark propagating through a QCD medium
these factors guarantee probability conservation. 
Indeed, the virtual corrections to the amplitude and the \cca, 
$2*(-\frac12)$, cancel the $+1$ coming from the scattering in the medium. 

It is convenient to introduce scaled momentum variables
\begin{equation}
  \label{eq:9}
  \k2/\mu = \U2\>, \quad \q2/\mu=\Q2\>, \quad \p2/\mu= \V2\>,
\end{equation}
where $\mu$ is an appropriate scale for the problem. 
In case the high energy fermion moves through a hot QCD plasma $\mu$
may be taken to be the inverse Debye screening length, while for cold
nuclear matter $\mu$ is naturally taken to be a typical transverse
momentum exchanged in a quark-nucleon scattering. In addition to the
factors $\T2_i$ and $F_i$ the graphs in Fig.~1
naturally include a quark-``particle'' cross section when
corresponding complex conjugate amplitudes are included. Thus, for
example, the graph of Fig.~1a
when multiplied by a similar complex conjugate amplitude, leads to the 
graph illustrated in Fig.~2a,
and this graph is clearly proportional to the quark-particle
scattering cross section in a two-gluon-exchange approximation.
We define a normalised quark-particle cross section\refup{r4} by 
\begin{equation}
  \label{eq:10}
  V(Q^2)= \frac1{\pi\,\sigma}\frac{d\sigma}{dQ^2}
\end{equation}
with
\begin{equation}
  \label{eq:11}
  \sigma = \int \frac{d\sigma}{d^2Q}\> d^2Q\>.
\end{equation}
In case the medium is cold nuclear matter the ``particle'' referred to 
above is a nucleon while in case the medium is a hot QCD plasma the
particle can be taken to be a quark or gluon.

In this paper we do not consider collisional energy loss. To ensure the
validity of the independent scattering picture and to guarantee that 
$d\sigma/dQ^2$ depend only on transverse momentum, it suffices to assume
that the energy transfer from the quark to a particle in the medium be
small compared to the incident energy\refup{r4}. 

It is useful to combine factors to define a Born-term gluon emission
amplitude $f_0(\U2,\V2)$ as
\begin{equation}
  \label{eq:12}
  \frac{N_c}{2C_F} \,f_0(\U2,\V2) \>=\> \frac1{C_F} \sum_{i=1}^g
\int d^2Q\> V(Q^2)\> \T2_i(\U2,\V2,\Q2)\> F_i\>.
\end{equation}
Here $f_0$ is an \underline{amplitude} for gluon emission although it also 
embodies colour factors and quark-particle scattering factors from the 
complex conjugate amplitude. 
The factor $N_c/2C_F$ is included to agree with our earlier choice of
normalisation\refup{r3} of $f_0(\U2,\V2)$.

 We note that the Born amplitude $f_0$ and the full
 amplitude $f$ to be introduced later in Sec.~\ref{sec:time}, as
 well as their impact-parameter images, $\tf_0(\B2)$ and
 $\tf(\B2)$, are two-dimensional vectors. 
 It is implied hereafter though we chose not to underscore $f$'s as
 transverse vectors.

It is convenient to change from momentum space to impact parameter
space. Since $f_0$ depends on the two momenta, $\U2$ and $\V2$, it
might be expected that two impact parameters would be
necessary. 
However, because $\k2$ and $\p2$ enter in \eqref{eq:7}  
only in the combination $\k2-x\p2$, the amplitude $f_0$ can depend 
on $\U2$ and $\V2$ only in the combination $\U2-x\V2$. Thus it is
possible to express $f_0$ in terms of a single ``impact parameter'' as
\begin{equation}
  \label{eq:13}
  f_0(\U2,\V2) \>=\> \int \frac{d^2B}{(2\pi)^2}\>
  e^{i\B2\cdot(\U2-x\V2)} \tilde{f}_0(\B2) 
\end{equation}
with
\begin{equation}
  \label{eq:14}
   \tilde{f}_0(\B2) \>=\> \int d^2(\U2-x\V2)\> e^{-i\B2\cdot(\U2-x\V2)}
 f_0(\U2,\V2)\>.
\end{equation}
Using 
\begin{equation}
  \label{eq:15}
  \frac{\U2}{U^2} \>=\> -\frac{i}{2\pi} \int d^2B\> e^{i\B2\cdot\U2} 
\>\frac{\B2}{B^2}\>,
\end{equation}
it is straightforward to find
\begin{equation}
  \label{eq:16}
   \tilde{f}_0(\B2) = -2\pi i \frac{\{\B2\}}{B^2} 
  \left(\left[\, 1-\tilde{V}(B(1-x))\,\right]+\left[\,1-\tilde{V}(B) \,\right]
- \frac{1}{N_c^2} \left[\, 1-\tilde{V}(-Bx)\,\right]\right),
\end{equation}
with 
$$
   \tilde{V}(B)\>=\> \int d^2Q\> e^{-i\Q2\cdot\B2}\> V(Q^2)\>; 
\qquad \tV(0)=1\>.
$$
The $\{\>\}$ symbol in \eqref{eq:16} is defined in \eqref{eq:2}.

Now let us compare \eqref{eq:16} with our previous small-$x$ results
in (4.5) and (4.25) of Ref.~\cite{r3}. 
Of the three terms on the right hand side of \eqref{eq:16} the first and
third terms come from the production terms in the medium, graphs a--c
of Fig.~1,
while the second term comes from the elastic scattering terms in the
medium, graphs d--g of Fig.~1.
In the small-$x$ limit the first and second terms on the right hand
side of \eqref{eq:16} become equal while the third term
vanishes. Comparing with (4.25) of Ref.~\cite{r3} we note that our
present result is larger than what we previously found by a factor of
2 in the small-$x$ limit. 
This factor of 2 is due to an incomplete treatment of virtual contributions 
in our earlier work.

\subsection{The Born amplitude for the hard scattering case}

In case the high energy quark is produced in the medium through a hard 
scattering one must also include the contribution coming when the
endpoint of the integration of the gluon emission coincides with the
time of the hard scattering. It is not important to know the details
of the hard scattering since we are here only interested in the
radiative gluon spectrum accompanying the hard scattering and
subsequent rescatterings in the medium. Thus we may imagine that the
gluon is produced in a collision of a highly virtual photon with a
quark in the medium, which collision transfers a large energy to the
struck quark which then rescatters in the medium and radiates a gluon.
\begin{center}\mbox{
\epsfig{file=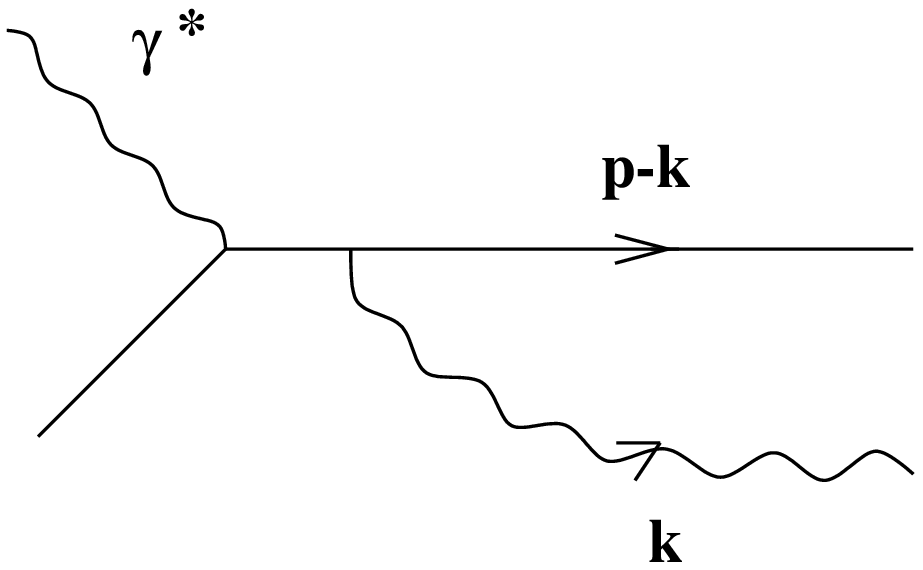,height=4cm,width=6cm}}\end{center}
%
The gluon emission amplitude is as given by $\T2_a$ in \eqref{eq:7a}.
Since the gluon is emitted after the hard scattering there is no
overall colour factor so that the basic Born term here is
\bminiG{eq:17}
\label{eq:17a} 
\frac{\{\U2-x\V2\}}{(\U2-x\V2)^2}\>, 
\eeeq
or
\beeq\label{eq:17b}
-2\pi i \frac{\{\B2\}}{B^2}
\emini
in impact parameter space. The expression given in \eqref{eq:17b}
must be added to $\frac{N_c}{2C_F}\tilde{f}_0(\B2)$, the Fourier
transform of the left hand side of \eqref{eq:12}.

\section{The time evolution of gluons in the medium \label{sec:time}}

After the gluon is emitted from the high energy quark, the quark-gluon 
system moves through the medium and carries out multiple 
scatterings\refupd{r9}{r10} with the particles of the medium.
It is not certain that the gluon will be produced as a physical gluon
until there is gluon emission in the complex conjugate
amplitude. Thus, we must follow the time evolution of the quark-gluon 
system in the amplitude up to the time of emission in the complex
conjugate amplitude. 
This is what will be done in this section. 

Let $f(\U2,\V2,t)$ be the quark-gluon amplitude at time $t$ starting
from $f_0(\U2,\V2)$ at $t=0$. $\U2$ is the scaled gluon momentum while 
$\V2-\U2$ is the scaled quark momentum as given in \eqref{eq:9} for
the graphs of Fig.~1.
The amplitude $f(\U2,\V2,t)$ is illustrated below.
\begin{center}\mbox{
\epsfig{file=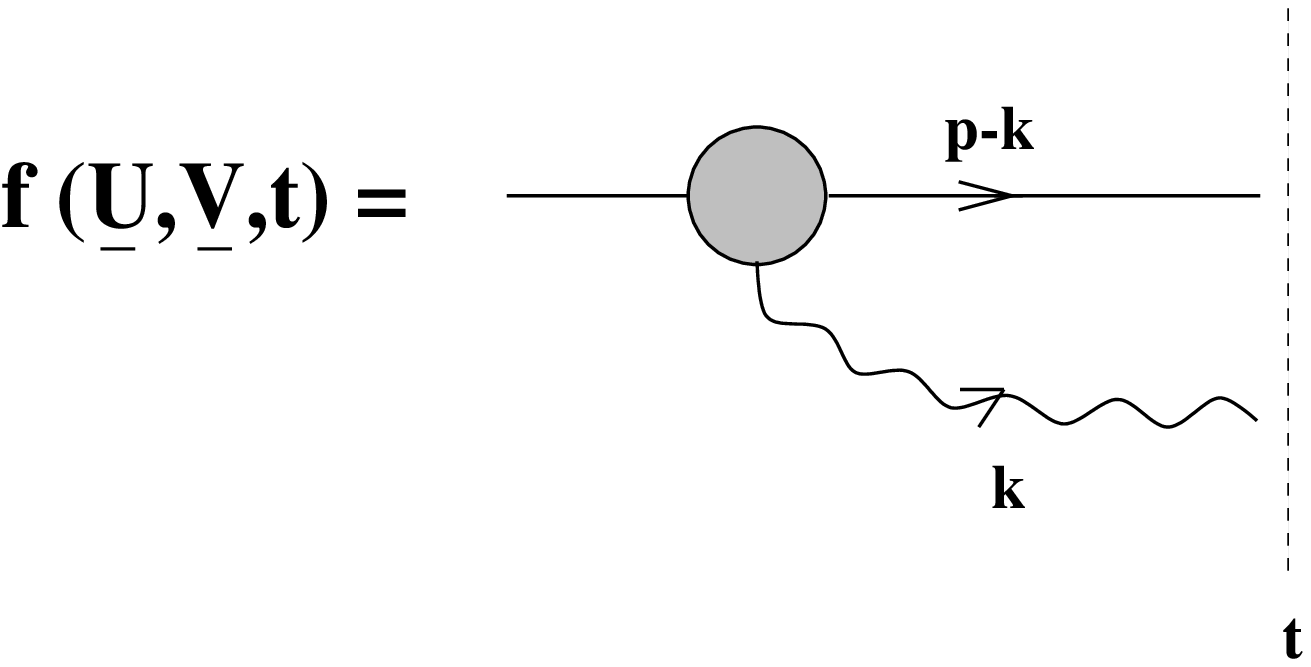,height=4cm,width=8cm}}\end{center}
We recall that $t$ is less than the time $t'$ at which the gluon is
emitted in the complex conjugate amplitude. Then the time-dependence
of $f$ comes partly from the free evolution of the quark-gluon system
and partly from interactions in the medium. Suppose $t_1$ is the time 
of the last interaction with the medium before time $t$. Although $f$ 
represents the gluon amplitude it also includes interactions of the
quark with the medium in the complex conjugate amplitude as already
noted for $f_0$. Then the possible interactions at $t_1$ are shown in
Fig.~3.
For example the graph shown in Fig.~3a
illustrates an inelastic interaction of the gluon with a particle in
the medium in the amplitude and an interaction of the quark with the
same particle in the complex conjugate amplitude. Perhaps the only
unusual graph in Fig.~3
is graph e where the only interaction is a forward elastic scattering
of the quark with a particle of the medium in the complex conjugate
amplitude. 

The amplitude $f(\U2,\V2,t)$ obeys the integral evolution equation
\begin{equation} \label{eq:18p}
\begin{split}
& f(\U2,\V2,t) = e^{i\phi (t-t_1)}f_0(\U2,\V2) 
+ \int_{t_1}^t dt'\, e^{i\phi (t-t')} 
\frac{\rho\sigma}{C_F}\int d^2Q\,V(Q^2)\left[
\frac{N_c}{2}f(\U2\!-\!\Q2,\V2\!-\!\Q2,t') \right. \\
&\left. 
-\frac1{2N_c}f(\U2,\V2\!-\!\Q2,t') - \frac{N_c}{2}f(\U2,\V2,t') 
-\frac{C_F}{2}f(\U2,\V2,t')-\frac{C_F}{2}f(\U2,\V2,t')
+ \frac{N_c}{2}f(\U2\!-\!\Q2,\V2,t')\right].
\end{split}
\end{equation}
The first term in the right hand side of \eqref{eq:18p} gives the free
evolution of the quark-gluon system between $t_1$ and $t$ in the
amplitude and of the quark in the \cca\ 
according to the phase factor 
\begin{equation}
  \label{eq:19}
  e^{i\phi(t-t_1)}\qquad\mbox{with}\quad 
  \phi= \frac{\k2^2}{2k}+\frac{(\p2-\k2)^2}{2(p-k)}-\frac{\p2^2}{2p}
  = \frac{(\U2-x\V2)^2\mu^2}{2x(1-x)p}\>.
\end{equation}
Differentiating over $t$ we obtain the equation
\begin{equation} \label{eq:18}
\begin{split}
  \frac{\partial}{\partial t}& f(\U2,\V2,t) =
\frac{i(\U2-x\V2)^2\mu^2}{2x(1-x)p} f(\U2,\V2,t)
+ \frac{\rho\sigma}{C_F}\int d^2Q\,V(Q^2)\left[
\frac{N_c}{2}f(\U2-\Q2,\V2-\Q2,t) \right. \\
&\left. 
-\frac1{2N_c}f(\U2,\V2-\Q2,t) - \frac{N_c}{2}f(\U2,\V2,t) 
-\frac{C_F}{2}f(\U2,\V2,t)-\frac{C_F}{2}f(\U2,\V2,t)
+ \frac{N_c}{2}f(\U2-\Q2,\V2,t)\right].
\end{split}
\end{equation}
The first term in the right hand side of \eqref{eq:18} comes from free
propagation and the rest of 
the terms come from the upper limit of the integration over $t_1$ at $t_1=t$. 
In \eqref{eq:18p}, \eqref{eq:18} $\rho$ is the density of
scatterers in the medium while $\sigma$ is the cross section of a
quark with a particle in the medium. 
The factor $V(Q^2)\sigma$
corresponds to the differential cross section for scattering the high
energy quark with momentum transfer $Q\mu$ while the $1/C_F$ factor
takes out the quark colour factor for quark scattering with a particle 
of the medium. 
The correct colour factors are then inserted in the various six terms
in the integrand in \eqref{eq:18}.
The six terms in the integrand in \eqref{eq:18} correspond, in order,
to the six graphs of Fig.~3.
The colour factors are
\bminiG{eq:20}
&F_a\>=\> -F_c\>=\> F_f\>=\> \frac{N_c}{2}\>, \\
&F_d\>=\> F_e\>=\> - \frac{C_F}{2}\>, \\
& F_b \>=\> -\frac{1}{2N_c} 
\>.
\emini
As in \eqref{eq:8} we include $(-1)$ factors in $F$ for all virtual
terms while factors $1/2$ are included for the terms with both gluon
lines, $q$ and $-q$, hooking to the same high energy quark or gluon line.

It is easy to see that $f(\U2,\V2,t)=f(\U2\!-\!x\V2,t)$ 
since the explicit factors of $\U2$ and $\V2$ in \eqref{eq:18} occur
in the combination $\U2-x\V2$ while $f(\U2,\V2,0)= f_0(\U2\!-\!x\V2)$. 
Introducing\refup{r3} 
\begin{equation}
  \label{eq:21}
  \tau= \frac{t}{\lambda}\,\frac{N_c}{2C_F}\>, \quad 
  \tk= \frac{\lambda\mu^2}{2x(1-x)p}\,\frac{2C_F}{N_c}
  = \frac{\lambda\mu^2}{2\omega(1-x)}\,\frac{2C_F}{N_c}
\end{equation}
with $\rho\sigma=1/\lambda$, where $\lambda$ is the mean free path of
the high energy quark in the medium, one can write \eqref{eq:18} as
\begin{equation}
  \label{eq:22}
\begin{split}
  \frac{\partial}{\partial \tau} &f(\U2-x\V2,\tau) =
i\tk(\U2-x\V2)^2 f(\U2-x\V2,\tau) \\
& \left.- \int d^2Q\,V(Q^2)\>\>\right(
\left[f(\U2-x\V2,\tau)- f(\U2-x\V2-(1-x)\Q2,\tau)\right]  \\
& \left. +\left[f(\U2-x\V2,\tau)- f(\U2-x\V2-\Q2,\tau)\right]
- \frac1{N_c^2}\left[f(\U2-x\V2,\tau)- f(\U2-x\V2+x\Q2,\tau)\right]\right).
\end{split}
\end{equation}
Now it is a simple matter to go to impact parameter space,
\begin{equation}
  \label{eq:23}
  \tilde{f}(\B2,\tau) \>=\> \int d^2(\U2-x\V2)\>
  e^{-i\B2\cdot(\U2-x\V2)} f(\U2-x\V2,\tau)\>,
\end{equation}
to find
\begin{equation}
  \label{eq:24}
\begin{split}
   \frac{\partial}{\partial \tau} \tf(\B2,\tau) &=
-i\tk\nabla^2_B \tf(\B2,\tau) \\
&-\left(\left[1-\tV(B(1-x)) \right] + \left[1-\tV(B) \right] 
- \frac1{N_c^2}\left[1-\tV(-Bx) \right]\right)\tf(\B2,\tau) \>.
\end{split}
\end{equation}
This is the basic Schr\"odinger-type evolution equation for the
propagation of the quark-gluon system in a QCD medium. It should be
solved with the initial condition 
\begin{equation}
\label{eq:incon}
  \tf(\B2,0)\>=\> \tf_0(\B2)\>,
\end{equation}
with $\tf_0(\B2)$ given in \eqref{eq:16}.

Comparing \eqref{eq:24} to (4.23) of Ref.~\cite{r3} we again find a
factor 2 discrepancy in the small-$x$ limit of the second term on the
right hand side of \eqref{eq:24} as compared to (4.23). 
Our previous calculation effectively amounted to keeping graphs a, b,
d and e of Fig.~3.
(Although graphs d and e were not explicitly 
considered, they were effectively included through the use of a mean
free path term, the $(-1)$ in the first term on the right hand side of
(4.16) of Ref.~\cite{r3}.)
That calculation was in error because the virtual terms, graphs c, d,
e, f of Fig.~3
are not completely taken into account in the mean free path treatment
of Ref.~\cite{r3}. We note that the coefficient of
$\tf(\B2,\tau)$, in the second 
line
of \eqref{eq:24}, the potential term, 
is of the same form as the three-body cross section used 
by Zakharov in eq.~23 of Ref.~\cite{r5}.

\section{The Born term for the \cca}

Now that we have calculated gluon emission in the amplitude and
followed its evolution in time, it becomes necessary to calculate
gluon emission in the \cca. Suppose the gluon is emitted at time $t$
in the \cca. As usual we integrate this emission time between elastic
or inelastic interactions with the medium. Except for colour factors
the calculation proceeds exactly as in Sec.~2. 
Suppose $t_{1}$, the time of interaction in the \cca\ 
with the particle in the medium, serves as the endpoint of the
$t$-integration. 
At $t_1$ the amplitude consists of a quark-gluon system described by
$f(t_1)$. The graphs describing the emission are shown in Fig.~4
where the vertical line indicates that the gluon is put on-shell. 
Terms to the left of the vertical line belong to the amplitude while
to the right belong to the \cca. We have rearranged the momenta so that
the gluon emission amplitude $f(\U2-x\V2,t_1)$ appears uniformly in
all the graphs. 

The ``colour factors'' for gluon emission in the \cca\ are
\bminiG{eq:25}
& \frac12\, F_a \>=\> -F_b \>=\> F_c \>=\> -F_d \>=\> -F_h \>=\> F_i 
\>=\> \frac{N_c}{2}\>, \\
&  F_e \>=\> -2\,F_g \>=\> -2\,F_j \>=\> C_F\>, \\
& F_f \>=\> -\frac1{2N_c}\>,
\emini
where again a factor of $(-1)$ for virtual terms and a factor of $1/2$
for identical particles have been included in \eqref{eq:25}. 
The gluon emission terms, analogous to \eqref{eq:7}, for the \cca\ are
\bminiG{eq:26}
   \label{eq:26a}
  \T2_a^* &=& \T2_e^*= \T2_g^* =\T2_h^*=-\T2_j^* 
\>=\>- \frac{\{\k2-x\p2\}^*}{(\k2-x\p2)^2}\>, \\
\T2_b^* &=& \T2_d^* =\T2_i^* 
= -\frac{\{\k2-x\p2-\q2\}^*}{(\k2-x\p2-\q2)^2} \>, \\
\T2_c^* &=& \frac{\{\k2-x\p2-(1-x)\q2\}^*}{(\k2-x\p2-(1-x)\q2)^2} \>, \\
\T2_f^* &=& \frac{\{\k2-x\p2+x\q2\}^*}{(\k2-x\p2+x\q2)^2} \>.
\emini
It is now straightforward to check that 
\begin{equation}
  \label{eq:27p}
\frac{1}{C_F}\sum_{i=a}^j
 \int d^2Q\, V(Q^2) \> T_i^*(\U2,\V2,\Q2)\>F_i
\>=\> -\frac{N_c}{2C_F} \,f_0^*(\U2-x\V2)\>,
\end{equation}
which, except for a {\em minus sign}, is similar to \eqref{eq:12}.
Taking the Fourier transform gives
\begin{equation}
  \label{eq:27}
  \int d^2(\U2-x\V2)\> e^{-i\B2\cdot(\U2-x\V2)} \frac1{C_F}
  \sum_{i=a}^j \int d^2Q\, V(Q^2)\> \T2_i^*(\U2,\V2,\Q2)\,F_i 
=  {\tf_0}^*(\B2)\,\frac{N_c}{2C_F}\>,
\end{equation}
with ${\tf_0}$ exactly the same function as given in \eqref{eq:16}.

Comparing with what was found in Ref.~\cite{r3}, it is easy to verify
that graphs a+b+c of Fig.~4
generate what was called the ``$Y$'' term and that this contribution is
equal to the first and second of the three terms 
   on the right hand side of \eqref{eq:16}. 
Graphs d+e+f generate what was called the ``$H$'' term which is equal to
the second and third terms 
   on the right hand side of \eqref{eq:16}. 
The virtual graphs, graphs g+h+i+j of Fig.~4
were not included in Ref.~\cite{r3} and it is easy to check using
\eqref{eq:25} and \eqref{eq:26} that they give $(-1)$ times the second term 
   on the right hand side of \eqref{eq:16}.
Thus in the small-$x$ limit where the third term 
on the right hand side of \eqref{eq:16}
is small we now find a factor 2, instead of the 3 coming from the sum of
the $Y+H$ graphs in Ref.~\cite{r3}, 
due to inclusion of the virtual terms.

\section{The spectrum of radiated gluons and energy loss}

In this section we remind the reader of the formula for the induced
spectrum of radiated gluons\refup{r3}. 
We then solve for the spectrum in the soft gluon limit for a volume of 
matter large enough that a high energy quark carries out many
scatterings. Finally we integrate the spectrum of radiated gluons to
find the radiative energy loss in circumstances where the energy loss
problem is dominated by soft gluons, that is when a gluon having
coherent length on the order of the dimensions of the medium has a
longitudinal momentum much less that that of the high energy quark. 

\subsection{The formula for the induced gluon radiation}

We give the formula for the induced gluon spectrum coming from a high
energy quark produced in a hard collision in the medium. Then the Born 
term in the amplitude will be associated either with a scattering in
the medium, as in \eqref{eq:12}, or with the hard vertex, as in
\eqref{eq:17a}. In case the gluon is radiated from a high energy
``on-shell'' quark entering the medium one simply drops the term
associated with the hard vertex. The induced gluon spectrum is
\begin{equation}
  \label{eq:28}
\begin{split}
  \frac{\omega dI}{d\omega\,dz} =& \frac{\as C_F}{\pi^2\,L}
\,2 \Re\int d^2U \left\{ \int_0^L\!\! dt_2\int_0^{t_2}\!\!dt_1\> 
 \rho\sigma  \frac{N_C}{2C_F}\, f(\U2\!-\!x\V2, t_2\!-\!t_1)
\cdot 
\rho\sigma \frac{N_C}{2C_F}\, f_0^*(\U2\!-\!x\V2)
\right.\\
&\left.
+  \int_0^L dt \>
\rho\sigma \frac{N_C}{2C_F}\, f(\U2-x\V2, t)
\cdot \frac{\{\U2-x\V2\}^*}{(\U2-x\V2)^2} \right\}_{\tk}^{\tk=0} \>.
\end{split}
\end{equation}
The various terms in \eqref{eq:28} have simple interpretations. 
The $\as C_F/\pi^2$ is the coupling of a gluon to a quark. The $1/L$
comes because we calculate the spectrum per unit length of the
medium. 
The factor 
$$
\frac{N_c}{2C_F}\, f(\U2-x\V2, t_2\!-\!t_1)\>\rho\sigma\> dt_1
$$ 
gives the number of scatterers in the medium, $\rho\sigma\> dt_1$,
times the gluon emission amplitude at $t_1$, then evolved to
$t_2$. The factor  
$$
-\frac{N_c}{2C_F}\, f_0^*(\U2-x\V2)\>\rho\sigma\> dt_2
$$ 
gives the number of scatterers times gluon emission in the \cca. 
The overall normalisation in the small-$x$ limit is fixed by comparing
with Ref.~\cite{r3}. We note that this normalisation is correct even
when $x$ is not small because the $\sqrt{1-x}$ present in the basic
amplitude \eqref{eq:6} but not included in our definition of $f_0$
cancels with a $1/1-x$ quark phase space factor. 

The second term on the right hand side of \eqref{eq:28} gives the
contribution of gluon emission due to the hard scattering in the
medium. If there is no hard scattering in the medium this term should
not be included.
We assume the hard scattering happens at $t=0$. 
This means that in the case the quark is produced by a hard scattering
in a medium $L$ is the length of matter measured from the production
point. 

The subtraction of the value of the integrals at
$\tk=0$ in \eqref{eq:28} eliminates the
medium-independent factorisation contribution\refupd{r1}{r2}. 

It is straightforward to simplify \eqref{eq:28}. Using
$\rho\sigma=1/\lambda$ and \eqref{eq:21} to write 
$dt=\frac{2C_F}{N_C}\lambda\,d\tau$, and defining 
$\tau_0=N_cL/2C_F\lambda$, one finds
\begin{equation}
  \label{eq:29}
\begin{split}
  \frac{\omega dI}{d\omega\,dz} =& \frac{\as N_c}{\pi^2\,\lambda}
\, \Re \int d^2U
\left\{ \int_0^{\tau_0}\!\! d\tau\,  \left(1-\frac{\tau}{\tau_0}\right)
f(\U2\!-\!x\V2,\tau)
\cdot f_0^*(\U2\!-\!x\V2) \right.\\
&\left.
+  \frac{1}{\tau_0}\int_0^{\tau_0} d\tau \> f(\U2-x\V2, \tau)
\cdot \frac{\{\U2-x\V2\}^*}{(\U2-x\V2)^2} \>\>\right\}_{\tk}^{\tk=0} \>.
\end{split}   
\end{equation}
It is convenient to express the integrals in impact parameter
space. Using \eqref{eq:13} and \eqref{eq:17b} one finds
\begin{equation}
 \label{eq:30}
\begin{split}
  \frac{\omega dI}{d\omega\,dz} =&
  \frac{\as N_c}{2\pi^3\,\lambda}\Re \left\{ \int_0^{\tau_0}\!\!
    d\tau\, 
 \left(1-\frac{\tau}{\tau_0}\right) {\tf}(\B2,\tau) 
\cdot \tf_0^*(\B2) \,\frac{d^2B}{2\pi}  \right.\\
&\left.
+ \frac{2\pi i}{\tau_0}\int_0^{\tau_0} d\tau \> {\tf}(\B2, \tau)
\cdot \frac{\{\B2\}^*}{B^2}\>\frac{d^2B}{2\pi}\>\> \right\}_{\tk}^{\tk=0} \>.
\end{split} 
\end{equation}
In the small-$x$ limit \eqref{eq:30} reduces to
\begin{equation}
  \label{eq:31}
\begin{split}
  \frac{\omega dI}{d\omega\,dz} =&
  \frac{2\as N_c}{\pi^2\,\lambda}\Re\>i \left\{ \int_0^{\tau_0}\!\!
    d\tau\, 
 \left(1-\frac{\tau}{\tau_0}\right) \frac{1-\tV(B^2)}{B^2} 
\B2\cdot \tf(\B2,\tau) 
 \,\frac{d^2B}{2\pi}  \right.\\
&\left.
+ \frac{1}{2\tau_0}\int_0^{\tau_0} d\tau \> 
\frac{\B2}{B^2}\cdot \tf(\B2, \tau)
\>\frac{d^2B}{2\pi}\>\> \right\}_{\tk}^{\tk=0} \>.
\end{split} 
\end{equation}
Comparing with (4.24) of Ref.~\cite{r3} we see that the first term on
the right hand side of \eqref{eq:31} is $2/3$ times (4.24). 
This $2/3$ is exactly the $2/3$ factor we discussed at the end of
Sec.~4. 

The second term on the right hand side of \eqref{eq:31}, due
to the hard scattering in the medium, was missed in Ref.~\cite{r3}. 
At first glance it might seem that this term is small compared to
the first term because of the $1/\tau_0$ in front of the
integral. However, the second term is enhanced by a $1/B^2$ compared
to the first term and $1/B^2$ is of the order of $\tau_0$ in the
dominant part of the integral\refup{r3}. 

\subsection{The induced spectrum}
\label{sec:specandloss}
It is not difficult to solve \eqref{eq:24}, in a logarithmic
approximation for small $B^2$, and to use that solution in
\eqref{eq:30} to obtain the induced gluon spectrum. The details of the 
procedure are given in Ref.~\cite{r3}, and here we emphasise the
differences which occur when $x$ is not necessarily small. 
When $B^2$ is small, and this will be the case for matter long enough
so that many scatterings occur, it is convenient to write the
``potential'' in \eqref{eq:24} as
\begin{equation}
  \label{eq:32}
  \left[1-\tV(B(1-x)) \right] + \left[1-\tV(B) \right] 
- \frac1{N_c^2}\left[1-\tV(-Bx) \right] \>=\> \frac14\,B^2\,\tu(B^2,x)\>.
\end{equation}
In terms of $\tv(B^2)$ used in Ref.~\cite{r3} one has
\begin{equation}
  \label{eq:33}
  \tu(B^2,x) \>=\> 2\left(1-x+\frac{C_F}{Nc}x^2\right) \tv(B^2)\>,
\end{equation}
where $\tv$, and $\tu$, have only logarithmic dependence on $B^2$ for
small $B^2$. In this small-$B$ limit $\tf_0(B)$, from \eqref{eq:16},
becomes 
\begin{equation}
  \label{eq:34}
  \tf_0(\B2) \>=\> -\frac{\pi i}{2}\, \tu(B^2,x)\> \{\B2\}\>.
\end{equation}
The only change from Ref.~\cite{r3} is $\B2\to\{\B2\}$ and
$\tv\to\tu$. Eqs.~(5.12) and (5.13) now become
\begin{equation}
  \label{eq:35}
  \tf(\B2,\tau) \>=\> -\frac{i\pi \{\B2\}}{2\cos^2\omega\tau}
  \exp\left(-\frac{i}{2}m\omega_0 B^2 \tan \omega_0\tau\right)
\end{equation}
and
\begin{equation}
  \label{eq:36}
   \tf(\B2,\tau) \>=\> \frac{2\pi i \{\B2\}}{B^2} \, 
  \frac{\partial}{\partial\tau}
  \exp\left(-\frac{i}{2}m\omega_0 B^2 \tan \omega_0\tau\right)  
\end{equation}
respectively. Parameters $m$ and $\omega_0$ are defined as before but
with $\tu$ replacing $\tv$. That is
\begin{equation}
  \label{eq:37}
  m = -\frac{1}{2\tk}\>, \quad \omega_0= \sqrt{i\tk\tu}\>, \quad 
  \tk= \frac{2C_F}{N_c}\,\frac{\lambda\mu^2}{2\omega(1\!-\!x)}\>. 
\end{equation}
Eqs.~\eqref{eq:35} and \eqref{eq:36} are two useful forms of the
solution to \eqref{eq:24} in the approximation of treating $\tu$ as a
constant, an approximation which should be good for small $B^2$. 

If one substitutes \eqref{eq:36} for $\tf(\B2,\tau)$ in the first term 
on the right hand side of \eqref{eq:30}, and \eqref{eq:35} for the
second term on the  right hand side of \eqref{eq:30}, then 
\begin{equation}
  \label{eq:38}
    \frac{\omega dI}{d\omega\,dz} =
  \frac{\as N_c}{\pi\lambda\tau_0}\left(1-x+\frac{x^2}{2}\right) 
\Re\left\{ \int_0^{\tau_0}\frac{d\tau}{\tau}\left[\, 
\left(\frac{\omega_0\tau}{\tan\omega_0\tau}-1\right) 
- \left(\frac{\omega_0\tau}{\sin\omega_0\tau\cos\omega_0\tau}-1\right)
\right]\right\} 
\end{equation}
emerges.
The two terms in the integrand in \eqref{eq:38} correspond exactly to
the terms on the right hand side of \eqref{eq:30}. In the small-$x$
limit the first term on the right hand side of \eqref{eq:38} is
smaller by a factor $1/3$ than (5.15) of Ref.~\cite{r3}, this factor
of $1/3$ being part of the $2/3$ found in Sec.~4, with the $2$ in the
$2/3$ going into changing a $\tv$ to a $\tu$. 

The $\tau$-integral in \eqref{eq:38} is easily done to give 
\bminiG{eq:39}
\label{eq:39a}
     \frac{\omega dI}{d\omega\,dz} =
  \frac{2\as C_F}{\pi\,L}\left[\,1-x+\frac{x^2}{2}\,\right] 
 \left( \ln\abs{\frac{\sin\omega_0\tau_0}{\omega_0\tau_0}}
   - \ln\abs{\frac{\tan\omega_0\tau_0}{\omega_0\tau_0}}\right),
\eeeq
or
\beeq
\label{eq:39b}
     \frac{\omega dI}{d\omega\,dz} =
  \frac{2\as C_F}{\pi\,L}\left[\,1-x+\frac{x^2}{2}\,\right] 
\ln\abs{\cos\omega_0\tau_0}\>.
\emini 
We remind the reader that \eqref{eq:39b} corresponds to the quark being
produced in the medium. For quark approaching the medium from outside
the spectrum is given by the first term in the right hand side of
\eqref{eq:39a}. 

Now we are in a position to generalise the gluon energy spectra \eqref{eq:39} 
for the case when the projectile is not a spin-1/2 fermion but a vector or
a scalar object. 
To this end we note that the basic gluon emission vertex
\eqref{eq:1} becomes diagonal in the gluon helicity basis, $s=\pm1$, 
  \begin{equation}
  \label{eq:ad1}
 \{v\}\cdot\eps^{(s)*} \>=\> \left(1-\frac{x}{2}\right)+s\frac{x}2 =
\left\{ \begin{array}{ll} 
1 & \mbox{for} \>\> s=+1\>, \\  
1-x & \mbox{for} \>\> s=-1\>.
\end{array}  \right.
  \end{equation}
In general, the structure of the gluon radiation vertex is\refup{rBL}  
  \begin{equation}
  \label{eq:ad2}
  \frac{2}{x(1-x)} \cdot \left[\, (1-x)^{\abs{s-rJ_P}}\, x^{\abs{r-r'}J_P}
  \, \right] ,
  \end{equation}
where $J_P$ is spin, and $r,r'=\pm1$ helicity states of the projectile 
before and after gluon emission. 
For the quark case we have $J_P=\frac12$ while helicity conserves,  
$r=r'$. 
Taken together with \eqref{eq:ad1}, this brings us back to \eqref{eq:1}:
$$
 \tilde{u}_{r'}(p-k)\>\gamma\cdot\eps^{(s)*}\> u_r(p) 
= \frac{2\delta_{rr'}}{x(1-x)}  \left[\, (1-x)^{1/2}\cdot\delta_{s,r}
+ (1-x)^{3/2}\cdot\delta_{s,-r} \, \right].
$$
Applying \eqref{eq:ad2} we obtain the expressions in the square brackets
in \eqref{eq:ad2} for the amplitude of gluon emission off a scalar
particle, $J_P=0$,
 \bminiG{eq:ad4}
  \left[\, (1-x)\cdot\left(\delta_{s,+1} + \delta_{s,-1}\right) \, \right], 
 \eeeq
and for the vector projectile (gluon), $J_P=1$, 
 \beeq\label{eq:ad5}
   \left[\, \left(1\cdot\delta_{r,s}+ (1-x)^2\cdot \delta_{r,-s}\right) 
   \delta_{r,r'} + x^2\cdot \delta_{r,s}\delta_{r,-r'}  \, \right]
 \emini
The last term in \eqref{eq:ad5} corresponds to helicity flip of the
incoming hard gluon. 
(The full answer is symmetric with respect to two
outgoing gluons, $x\leftrightarrow(1\!-\!x)$, $s\leftrightarrow r'$, 
as it should be.)

Adding together the squared helicity amplitudes we  
finally obtain the $x$-dependent factors $X_{J_P}$ 
in the gluon energy spectrum \eqref{eq:39}
for different projectiles: 
\bminiG{eq:ad6}
 X_{\frac12}(x)  &=& \left[\>\frac{1+(1-x)^2}{2}\>\right]
 = \frac{x}2\cdot \frac{1+(1-x)^2}{x} , \\
 X_{0}(x)  &=& \quad \left[\>1-x\>\right] \quad = \quad\frac{x}2\cdot 2(1-x), \\
 X_{1}(x) &=& \left[\>\frac{1+(1-x)^4+x^4}{2(1-x)}\>\right] 
 = \frac{x}2\cdot \left\{\frac{x}{1-x}+\frac{1-x}{x}+x(1-x)\right\}.
\emini
Factors \eqref{eq:ad6} are identical in the soft limit, $X_{J_P}(0)=1$. 
They are proportional to the corresponding parton splitting functions. 

If the projectile corresponds to colour representation, $R$, 
which is different from the fundamental representation, the colour factor $C_F$ 
in the gluon energy spectrum \eqref{eq:39} 
should be replaced by the appropriate quadratic Casimir operator, $C_R$.
The colour factor $1/N_c^2$ in the third term in the ``potential''
\eqref{eq:32} should be replaced by 
$$
    -\frac{1}{N_c^2}\>\Longrightarrow\> \frac{2C_R-N_c}{N_c}\>,
$$
which leads to the substitution $C_F\to C_R$ 
in the expression \eqref{eq:33} for $\tilde{u}$.

\subsection{The energy loss}

The energy loss per unit length,
$$
-\frac{dE}{dz} \>=\> \int_0^\infty  \frac{\omega dI}{d\omega\,dz}\>d\omega\>, 
$$
can be evaluated easily if the dominant values of 
$\omega\sim \mu^2L^2/\lambda$ are such that the small-$x$ approximation 
can be used when doing the $\omega$-integral of \eqref{eq:39}. 
In this case
\begin{equation}
  \label{eq:40}
  -\frac{dE}{dz} \>=\> \frac{\as\,N_c}{4}\>\frac{\mu^2\,L}{\lambda}
  \>\tv(\tau_0^{-1})\>,
\end{equation}
where we have used $\tu\simeq2\tv$ in the small-$x$ limit.
We remind the reader that in \eqref{eq:40} $L$ is the length of material
traversed by the quark beyond its production point. 

The relative contributions of the first and second terms on the right
hand side of \eqref{eq:38} to \eqref{eq:40} are $1/3$ and $2/3$
respectively.
Thus for a quark approaching the medium {\em from outside}\/
\eqref{eq:40} is replaced by
\begin{equation}
  \label{eq:40p}
  -\frac{dE}{dz} \>=\> \frac{\as\,N_c}{12}\>\frac{\mu^2\,L}{\lambda}
  \>\tv(\tau_0^{-1})\>,
\end{equation}
Comparing with Ref.~\cite{r3} in the small-$x$ limit we note that the
first term on the right hand side of \eqref{eq:39a} is a factor $1/3$
times the expression given in (5.16) of Ref.~\cite{r3} with, of
course, $\tv$ replaced by $\tu$ in the definition of $\omega_0$ in the 
present result. 

Comparing energy loss with jet broadening\refup{r4}, for example for a 
jet produced in matter, we find
\begin{equation}
  \label{eq:41}
   -\frac{dE}{dz} \>=\>  \frac{\as\,N_c}{4}\>p_{\perp W}^2\>,
\end{equation}
the coefficient being a factor of 2 {\em larger}\/ than that quoted in
Ref.~\cite{r4}.

\section{Comparing to the method of Zakharov}

Recently B. Zakharov\refupd{r5}{r6} has proposed a concise and elegant 
formulation for describing and calculating the energy loss of high
energy partons in hot and cold matter. At first sight Zakharov's
formalism appears very different from 
that of BDMPS. In Zakharov's picture radiative energy loss of a high
energy quark is described by the interaction of a high energy
colour-neutral quark-antiquark-gluon system with the QCD medium it
passes through. No trace of a quark radiating a gluon as it passes
through a medium and carries out multiple scatterings with that medium
remains visible. Nevertheless, as we shall see below the two formalisms 
are in fact exactly equivalent. 
We begin by casting \eqref{eq:30} in a form given by Zakharov. 
Then we shall attempt to explain how one can intuitively see the
equivalence of the two formalisms.

\subsection{Quantitative equivalence of the two formalisms}

In this section we shall show that \eqref{eq:30} can be expressed in a 
form identical to Eq.~4 of Ref.~\cite{r6}.
It is convenient to go back to the unscaled variables used in \eqref{eq:28}, 
and also to use
\begin{equation}
  \label{eq:42}
  \tf(\B2_2,t_2) \>=\> \int d^2B_1\> G(\B2_2,t_2;\B2_1,t_1)\> \tf_0(\B2_1)\>,
\end{equation}
where $G$ is exactly as given in (5.6) of Ref.~\cite{r3} with, of
course, the replacement of $\tv$ by $\tu$ being understood. Then
\eqref{eq:30}, or \eqref{eq:28}, takes the form
\begin{equation}
  \label{eq:43}
\begin{split}
    \frac{\omega dI}{d\omega\,dz} = \frac{\as C_F}{L\,\pi^3}
 \left(\frac{N_c}{2C_F}\right) \> \Re\> i
\int d^2B_2\int d^2B_1\>\frac{\{\B2_1\}}{B_1^2}
\cdot \left\{ \int_0^\infty dz_2 
  \rho(z_2)\sigma \int_0^\infty dz_1\> \tf_0^*(\B2_2) \right. \\
\left.
G(\B2_2,z_2;\B2_1,z_1)\frac{B_1^2\tu(B_1,x)}{4} \frac{N_c}{2C_F}\,
\rho(z_1)\sigma 
+ \int_0^\infty dz_2 \rho(z_2)\sigma\> \tf_0^*(\B2_2)G(\B2_2,z_2;\B2_1,0)
\right\}_{\tk}^{\tk=0} , 
\end{split}
\end{equation}
where we have allowed the integrations over $z_1$ and $z_2$ to go to
$\infty$. It is understood, however, that $\rho(z)$ vanishes outside
the interval $0\le z\le L$. We have used $z$ rather than $t$ as a
variable to emphasise the spatial dependence of the density $\rho$. In 
what follows it is not important that $\rho$ be uniform in the region 
$0< z< L$, as we are only going to the equations for $G$ and not the
explicit solution. 
The Green function $G$ obeys the same differential equation as given for
$\tf$ in \eqref{eq:24}. Using \eqref{eq:21} and \eqref{eq:32} one
finds 
\bminiG{eq:44}
  \label{eq:44a}
 \frac{\partial}{\partial z_2} G \>=\> 
-i\left(\frac{\tk\,N_c}{2\lambda\, C_F}\right) \nabla^2_{B_2} G 
- \frac14 B_2^2\tu\,\frac{N_c}{2C_F}\,\rho(z_2)\sigma\> G\>,
\eeeq
and
\beeq 
  \label{eq:44b}
 \frac{\partial}{\partial z_1} G \>=\> 
 i\left(\frac{\tk\,N_c}{2\lambda\, C_F}\right) \nabla^2_{B_1} G 
+ \frac14 B_1^2\tu\,\frac{N_c}{2C_F}\,\rho(z_1)\sigma\> G\>.
\emini
Using \eqref{eq:44b} we 
substitute the combination of $\partial G/\partial z_1$ and
$\nabla^2_{B_1}G$ for the 
$\frac14 B_1^2\tu\,\frac{N_c}{2C_F}\,\rho(z_1)\sigma\> G$ 
in the first term on the right hand side of \eqref{eq:43}. 
The lower limit of the $z_1$-integral of $\partial G/\partial z_1$ 
exactly cancels
the second term on the right hand side of \eqref{eq:43} 
while its upper limit, at $z_2$, contains 
$$
G(\B2_2,z_2;\B2_1,z_2) \>=\> \delta^2\left(\B2_2-\B2_1\right) ,
$$
comes out $\tk$-independent and therefore cancels due to 
the $\tk$ versus $\tk\!=\!0$ subtraction indicated in \eqref{eq:43}.

Thus, we arrive at
\begin{equation}
  \label{eq:45}
\begin{split}
    \frac{\omega dI}{d\omega\,dz} =& \frac{\as C_F}{L\,\pi^3}
 \left(\frac{\tk\,N_c}{2\lambda C_F}\right)\> \Re\left\{ 
 \int d^2B_2\int d^2B_1\>\frac{\{\B2_1\}}{B_1^2}\right.\\
&\left.
 \int_0^\infty dz_1 \int_{z_1}^\infty dz_2 
  \rho(z_2)\sigma \left(\frac{N_c}{2C_F}\right) 
\cdot \tf_0^*(\B2_2)\> \nabla^2_{B_1} G(\B2_2,z_2;\B2_1,z_1)
\right\}_{\tk}^{\tk=0} .
\end{split}
\end{equation}
Now, using \eqref{eq:34} and \eqref{eq:44a} gives
\begin{equation}
  \label{eq:46}
\begin{split}
 \frac{\omega dI}{d\omega\,dz} = -\frac{2\as C_F}{L\,\pi^2} 
 \left(\frac{\tk\,N_c}{2\lambda C_F}\right)\> 
&
\Re\>i
 \int\!\!\!\int d^2B_1\,d^2B_2\>\frac{\{\B2_1\}\cdot\{\B2_2\}}{B_1^2\,B_2^2}\\
\nabla^2_{B_1} \int_0^\infty dz_1 \int_{z_1}^\infty dz_2 
&\left\{ \frac{\partial}{\partial z_2} G(\B2_2,z_2;\B2_1,z_1)
+ i\left(\frac{\tk\,N_c}{2\lambda C_F}\right)\nabla^2_{B_2} 
G(\B2_2,z_2;\B2_1,z_1)\right\}_{\tk}^{\tk=0} .
\end{split}
\end{equation}
The $z_2$-integral of the $\partial G/\partial z_2$ term vanishes: 
at the upper limit, $z_2=\infty$ because $G(z_2\!=\!\infty)=0$, 
and at the lower limit due to $\tk$-subtraction, as before. 
The remaining term in \eqref{eq:46} can be simplified by writing
\begin{equation}
  \label{eq:47}
  \frac{\{\B2\}}{B^2} = \frac12\> \{\nabla_{B}\}\, \ln B^2\>.
\end{equation}
After integrating the $\{\nabla_{B}\}$ and $\nabla_B^2$ terms by parts 
one finds 
\begin{equation}
  \label{eq:48}
\begin{split}
  \frac{\omega dI}{d\omega\,dz} &= \frac{2\as C_F}{L\,\pi^2} 
 \left(\frac{\tk\,N_c}{2\lambda C_F}\right)^2\> 
\Re
 \int\!\!\!\int d^2B_1\,d^2B_2 
 \int_0^\infty dz_2 \int^{z_2}_0 dz_1 \\
&\left. \left[\, \nabla_{B_1}^2\,\frac12\ln B_1^2\,\right]\,
   \left[\, \nabla_{B_2}^2\,\frac12\ln B_2^2\,\right]
 \{\nabla_{B_1}\}\cdot\{\nabla_{B_2}\}^*  
G(\B2_2,z_2;\B2_1,z_1)\right|_{\tk}^{\tk=0} .
\end{split}
\end{equation}
Using 
\begin{equation}
  \label{eq:49}
   \{\nabla_{B_1}\}\cdot\{\nabla_{B_2}\}^* \>=\>
   \left(1-x+\frac{x^2}{2}\right)\nabla_{B_1}\cdot \nabla_{B_2}
\end{equation}
and
\begin{equation}
  \label{eq:50}
   \nabla_{B}^2\,\frac12\ln B^2 \>=\> 2\pi \delta^2(\B2)\>,
\end{equation}
one finds
\begin{equation}
  \label{eq:51}
 \frac{\omega dI}{d\omega\,dz} = \frac{2\as C_F}{L} 
\left[4\!-\!4x\!+\!2x^2\right] 
\left(\frac{\tk N_c}{2\lambda C_F}\right)^2
\left. \Re\!\! \int_0^\infty \!\!\!dz_2 \int^{z_2}_0 \!\!\!dz_1 
\left[\, \nabla_{B_1}\!\! \nabla_{B_2} G(\B2_2,z_2;\B2_1,z_1)
\,\right]_{B_1=B_2=0} \right|_{\tk}^{\tk=0} .
\end{equation}
Noting that 
$$
\frac{\tk\,N_c}{2\lambda C_F} \>=\> \frac{\mu^2}{2x(1-x)p}\>,
$$
and that $\mu$ normalises our ``impact parameter'' $B$ 
to physical transverse coordinate, $\xu$, so that 
$$
  \mu\, \nabla_{B_1}\>=\> \nabla_{\xu}\>, \qquad \>\> 
  \mu^2\, G(\B2_2,0;\B2_1,0) \>=\> \delta^2(\xu_2-\xu_1)\>,
$$
we see that \eqref{eq:51} takes exactly the same form as Eq.~4 in
Ref.~\cite{r6}. 

It is not immediately clear that the subtraction terms, done at
$\tk\!=\!0$ in our case and at zero matter density in Ref.~\cite{r6},
are identical. There is, however, a physical argument that shows that
they should be the same. 
The $\tk\!\to\!0$ limit is equivalent to the $\omega\!\to\!\infty$
limit. However, at large $\omega$ the gluon is surely radiated outside 
the medium and since the high energy quark is produced in
the medium, 
the gluon has no knowledge of the medium whatsoever. 
Thus, subtracting out the zero-density calculation is the same as
subtracting out the large-$\omega$ gluons. 

In closing we note\refup{r6} that the two terms on the right hand side 
of \eqref{eq:38} correspond to integrations when $z_1$ and $z_2$ are
in the medium and  when $z_1$ is inside the medium while $z_2$ is
outside, respectively. 
Thus a formula like \eqref{eq:51} but with  
$z_1$ and $z_2$ restricted to lie in the medium reproduces the induced 
radiation off a high energy quark approaching the medium from
outside. 
In this case the subtraction at $\tk\!=\!0$ subtracts out the
so-called factorisation contribution\refupd{r1}{r2}. 
It does not appear that this subtraction can be done in terms of a
zero density limit as the subtraction term has a (weak) matter dependence.

\subsection{Why the two formalisms are equivalent}

In this section we describe, very qualitatively, how the two
formalisms are related. We do this in the context of a hard scattering 
producing a high energy quark jet in a finite-size medium. 
The quark then radiates a gluon either in the medium of after it has
left the medium. In the BDMPS approach one calculates the gluon
emission amplitude and evolves it in time up to the time that the
gluon is also emitted in the \cca, at which time it is certain that
the gluon will be produced and is not just a virtual fluctuation. 
After the gluon is emitted in the \cca\
the gluon emission spectrum is determined and it is not
necessary to follow the system any further in time.

Suppose the gluon is emitted at $t=t_1$ in the amplitude and at $t_2$
in the \cca. In evolving the amplitude, and the \cca, between $t_1$
and $t_2$ the quantum mechanical phase depends on the energy of the
quark and of the gluon in the amplitude and on the quark in the \cca\ 
as indicated in \eqref{eq:19} where the phase contribution from the
\cca\
comes with a sign opposite to that of the amplitude. 
Thus, formally, one may insert the phase from the \cca\
into the amplitude by introducing a \underline{negative} kinetic term
in the effective 2--dimensional Lagrangian, 
and that is what is done in Refs.~\citd{r5}{r6} and \cite{r11}. 

As the quark-gluon system in the amplitude, and the quark in the \cca, 
evolve between $t_1$ and $t_2$ there may be inelastic collisions with
the medium involving both the amplitude and the \cca. 
If the elastic scattering amplitude of a quark with a particle in the
medium is purely imaginary then the total inelastic scattering cross
section is given by the forward elastic amplitude. 
Then the inelastic contribution can be taken into account, 
solely in the amplitude, 
if one brings the quark from the \cca\ 
to the amplitude as an antiquark\refupd{r5}{r6}. 

Thus one can consider the evolution of a quark-antiquark-gluon system 
in the amplitude, and with the antiquark
having a negative kinetic energy term, as equivalent to the evolution
of a quark-gluon system in the amplitude and a quark in the \cca. 
Forward elastic scatterings of the $q\bar{q}g$ system with particles in 
the medium are equivalent to the various elastic and inelastic
reactions of the original problem. 
These forward elastic scatterings may be viewed as a two-body
imaginary potential between the various pairs of the three-body
system. This is also apparent in \eqref{eq:22} and \eqref{eq:24}. 

Thus, formally, one can represent gluon emission in terms of the
evolution of the amplitude for the three-body 
quark-antiquark-gluon system in a medium. 
There is a bit of a mystery in both approaches, and that concerns the
number of independent transverse variables required to describe the
process. One might expect there to be two independent impact
parameters necessary to describe the three-body evolution. 
In the BDMPS approach one also would, in general, expect two impact
parameters to be necessary to describe the amplitude $f(\U2,\V2,\tau)$ 
since there would seem to be two independent momenta, $\U2$ and
$\V2$. In the  BDMPS approach we have seen that $\U2$ and
$\V2$ only appear in the combination $\U2-x\V2$, as is apparent from 
\eqref{eq:7},  \eqref{eq:22} and \eqref{eq:23}, so that
only one coordinate is required. 

The structure of the potential in \eqref{eq:24} can be visualised as
pairwise interactions between a quark in the 
amplitude put
at transverse coordinate $\underline{0}$,  the gluon at coordinate 
$\B2$ and the quark in the \cca\
at $x\B2$. 
This same result was earlier found by Zakharov in Ref.~\cite{r5}.
If $\xu_q$, $\xu_{\bar{q}}$ and $\xu_g$ are the transverse coordinates 
of the quark, the ``antiquark'' and the gluon then the two-body
interactions only depend on relative distances, so that the total
momentum ${\cal{P}}$ is conserved:
$$
  (1-x)\cdot\dot{\xu}_q + x\cdot \dot{\xu}_g - \dot{\xu}_{\bar{q}} 
\>=\>\underline{\cal{P}} = \mbox{const} \>.
$$ 
The factors $(1-x)$, $x$ and $(-1)$ correspond to the relative
``masses'' of the quark, gluon and antiquark, respectively.
With the boundary conditions 
${\xu}_g-{\xu}_{\bar{q}}=0={\xu}_q-{\xu}_{\bar{q}}$ 
at $t=t_1,t_2$ one gets
$$
    (1-x)\left({\xu}_q - {\xu}_{\bar{q}}\right) 
  \>+\> x\left({\xu}_g - {\xu}_{\bar{q}}\right)\>=\> 0\>.
$$
Choosing $\xu_q\equiv0$ one finds ${\xu}_{\bar{q}}=x\,{\xu}_q$ just as 
in the BDMPS approach. 

\vspace {1 cm}
\noindent
{\large\bf Acknowledgements}

\noindent
Three of the authors (BDM) thank the LPTHE for the kind hospitality
during the times when much of this work was done. 
We are grateful to Slava Zakharov for stimulating discussions, in
particular, about his approach to the energy loss problem,
and to Yuri Kovchegov for help.

\newpage


\vfill

\begin{center}\mbox{
\epsfig{file=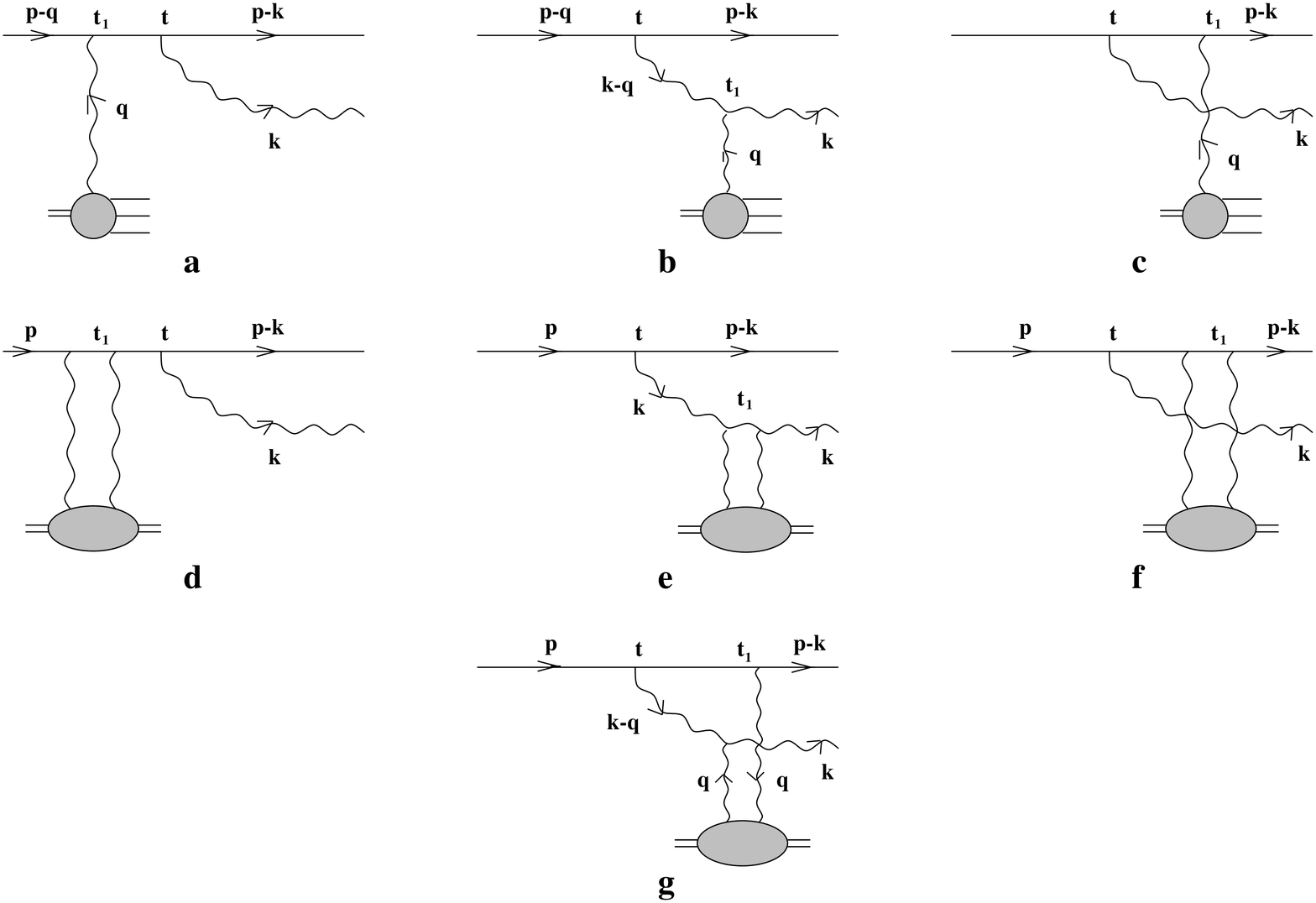,height=13.5cm,width=15cm}}\end{center}
\begin{center}Figure 1. 
\end{center}
\vfill

\newpage

\vfill

\begin{center}\mbox{
\epsfig{file=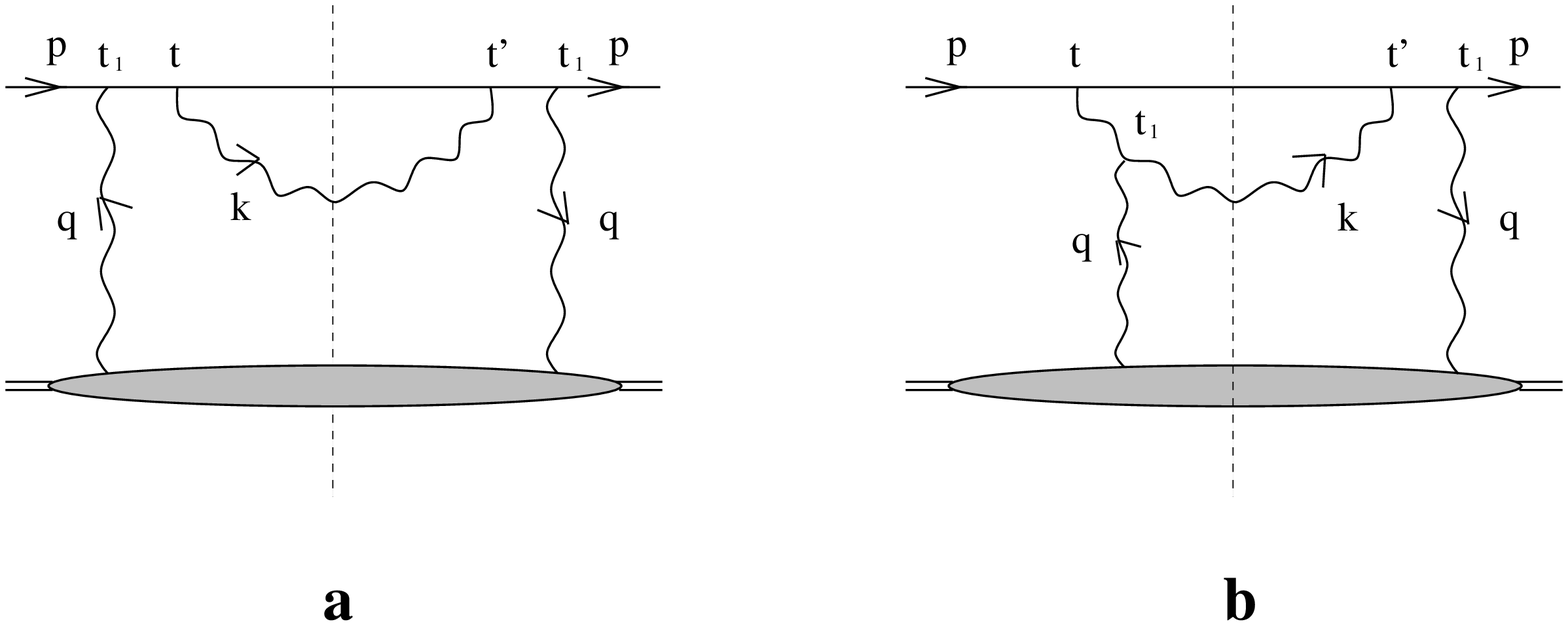,height=5.5cm,width=12cm}}\end{center}
\begin{center}Figure 2. 
\end{center}

\vfill


\begin{center}\mbox{
\epsfig{file=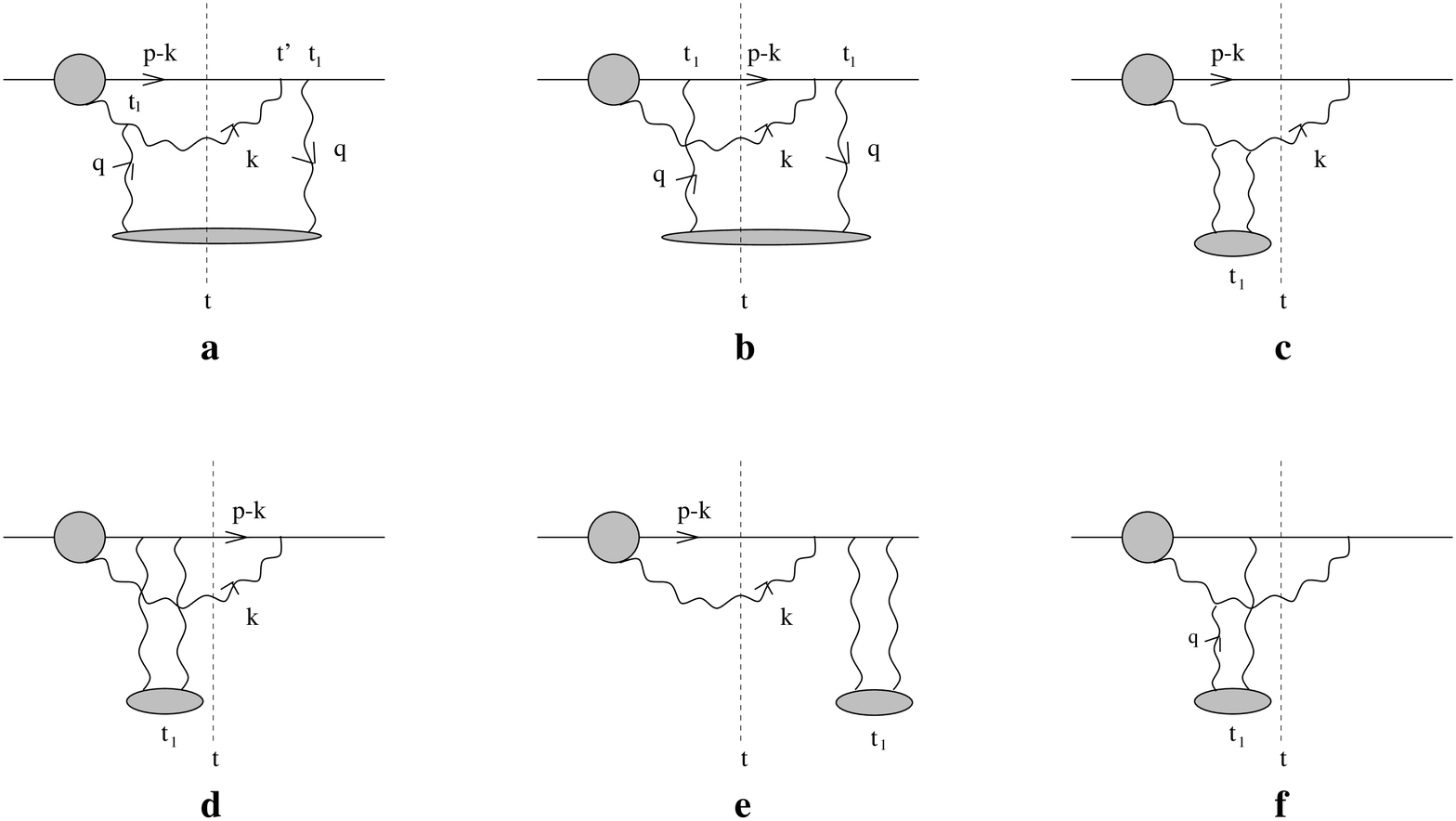,height=9cm,width=15cm}}\end{center}
\begin{center}Figure 3. 
\end{center}

\vfill

\newpage

\vfill

\begin{center}\mbox{
\epsfig{file=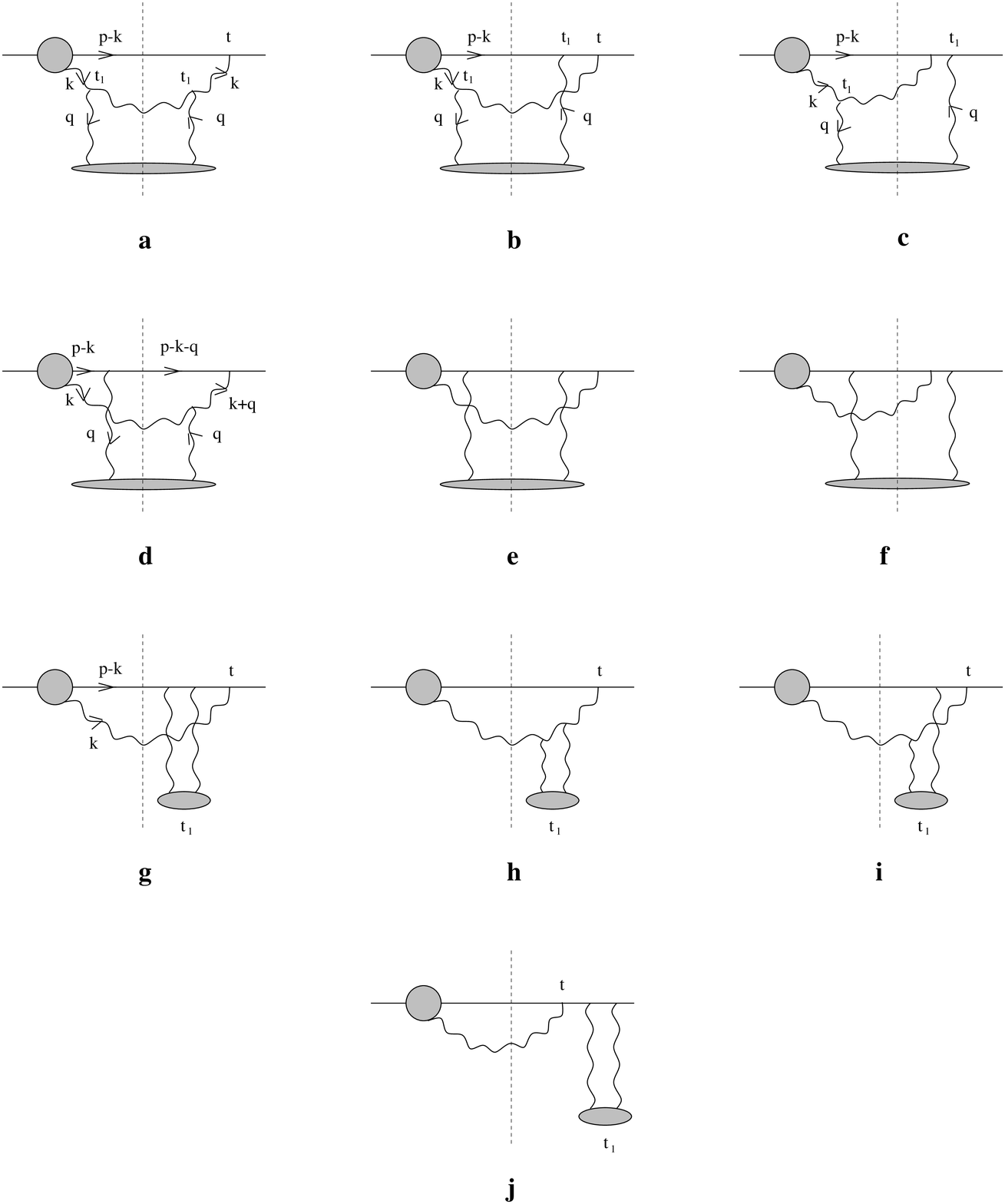,height=18cm,width=15cm}}\end{center}
\begin{center}Figure 4. 
\end{center}

\vfill

\end{document}